\documentclass[12pt]{iopart}
\usepackage{iopams}  
\usepackage{graphicx}
\usepackage{xcolor}
\newcommand{\cd}{c^{\dagger}}
\newcommand{\tR}{t_{1R}}
\newcommand{\tl}{t_{1L}}
\newcommand{\ee}{\rm{e}}
\newcommand{\ii}{\rm{i}}

\begin{document}
\title[]{Solvable non-Hermitian skin effects and real-space exceptional points: Non-Hermitian generalized Bloch theorem}
\author{Xintong Zhang, Xiaoxiao Song, Shubo Zhang, Tengfei Zhang, Yuanjie Liao, Xinyi Cai, Jing Li$^\dagger$}
\address{Department of Optical Science and Engineering, Shanghai Ultra-Precision Optical Manufacturing Engineering Center, Fudan University, Shanghai 200438, China}
\ead{$^\dagger$lijing@fudan.edu.cn}

\vspace{10pt}

\begin{indented}
\item[]April 2023
\end{indented}

\begin{abstract}
Non-Hermitian systems can exhibit extraordinary boundary behaviors, known as the non-Hermitian skin effects, where all the eigenstates are localized exponentially at one side of lattice model. To give a full understanding and control of non-Hermitian skin effects, we have developed the non-Hermitian generalized Bloch theorem to provide the analytical expression for all solvable eigenvalues and eigenstates, in which translation symmetry is broken due to the open boundary condition. By introducing the Vieta’s theorem for any polynomial equation with arbitrary degree, our approach is widely applicable for one-dimensional non-Hermitian tight-binding models. With the non-Hermitian generalized Bloch theorem, we can analyze the condition of existence or non-existence of the non-Hermitian skin effects at a mathematically rigorous level. Additionally, the non-Hermitian generalized Bloch theorem allows us to explore the real-space exceptional points. We also establish the connection between our approach and the generalized Brillouin zone method. To illustrate our main results, we examine two concrete examples including the Su-Schrieffer-Heeger chain model with long-range couplings, and the ladder model with non-reciprocal interaction. Our non-Hermitian generalized Bloch theorem provides an efficient way to analytically study various non-Hermitian phenomena in more general cases.

\end{abstract}
\maketitle

\section{Introduction}
Non-Hermitian physics has attracted significant studies since the discovery of parity-time symmetric systems which can provide entirely real spectrum without Hermiticity\cite{Bender1998,El-Ganainy2018}, and the sufficient and necessary condition of the real spectrum is related to the pseudo-Hermiticity and quasi-Hermiticity\cite{Mostafazadeh2002,Mostafazadeh2002b}. Actually, due to the non-equivalent energy exchange with the environment, the non-Hermitian Hamiltonian is useful to describe the realistic physical systems with complex spectrum, and the research on driven or dissipative open systems is on the rise in many fields, such as atomic and molecular physics\cite{Feshbach1958}, condensed matter physics\cite{Lindblad1976}, and mesoscopic physics\cite{Colbert1992}. Owing to the formal equivalence between wave equations and Schrödinger equation, many non-Hermitian classical systems have also been realized, such as photonics\cite{Zhu2020,Zhong2021,Chen2021,Wang2021,Parto2021}, electrical circuits\cite{Yoshida2020,Hofmann2020,Xu2021,Zhang2021,Zou2021,Kononchuk2022}, acoustics\cite{Jing2018,Acoustics2021}, and magnetics\cite{Flebus2020,Deng2022}. One of the most significant characteristics of non-Hermitian systems is the exceptional point\cite{Heiss2004,Bender2007}, where the degeneracy of eigenvalues always comes along with the degeneracy of eigenstates, beyond any known phenomenon from their Hermitian counterparts, thus providing the potential application in the enhancement of sensitivity\cite{Chen2017}. 

In recent years, the interplay between non-Hermitian physics and topological physics has revealed new phenomena in non-Hermitian systems\cite{Gong2018,Ghatak2019,Ashida2020,Bergholtz2021}. The traditional band topology can be traced back to the discovery of integer quantum Hall effect\cite{Klitzing1980,Thouless1982,Hatsugai1993}. For Hermitian systems, a nontrivial Bloch band invariant implies the emergence of topological edge states under the open boundary condition\cite{Bernevig2006,Schnyder2008}. However, similar bulk-boundary correspondence appears to be violated in non-Hermitian systems. On one hand, the spectrum under open boundary condition is far different from that under periodical boundary condition, particularly, the topological phase transition points are not equivalent to the bulk gap-closing points\cite{Lee2016,Yao2018}, and the traditional topological invariant cannot predict the topological edge states. On the other hand, such a non-consistency of spectra always comes along with a new phenomenon, called the non-Hermitian skin effects\cite{Yao2018,Alvarez2018,Okuma2020,Yokomizo2021,Kim2021}, where all the eigenstates for open systems are exponentially localized at the boundary of the lattice. Notably, the precondition of non-Hermitian skin effects relies on the fact that, unlike the Hermitian operators, the non-Hermitian operators only require the right eigenvector to be orthogonal to the left eigenvector\cite{Brody2013}. However, the non-Hermitian skin effects cannot be predicted by the conventional topological invariant.

Several efforts have been made to predict the non-Hermitian skin effects and restore the bulk-boundary correspondence in non-Hermitian systems. The generalized Brillouin zone (GBZ) method can predict the non-Hermitian skin effects by calculating the trajectory of the generalized Brillouin zone\cite{Yao2018,Yokomizo2019,Kawabata2020,Yang2020}. The transfer-matrix method can explain the connection between the non-Hermitian skin effects and the non-consistency of spectra by deriving the transfer matrix\cite{Kunst2019}. The biorthogonal bulk-boundary correspondence method is used to study the exactly solvable boundary modes in the lattice with destructive interference, and the biorthogonal polarization exhibits a jump when phase transition occur\cite{Kunst2018,Edvardsson2020}. Additionally, the exact solution approach has been proposed to predict the non-Hermitian skin effects by solving the analytical expression of eigenstates under open boundary conditions\cite{Guo2021,Guo2021b,Liu2021}. In this work, we focus more on the exact solution approach, because it can give us a direct understanding and full control of non-Hermitian skin effects. However, it is still challenging to extend the exact solution approach to be applicable for more complicated models, due to the difficulty of solving polynomial equations with a degree proportional to the hopping range\cite{Edvardsson2022}.

To address this challenge, we improve the existing exact solution approach by introducing Vieta's theorem for any general polynomials with arbitrary degree. We call this approach the non-Hermitian generalized Bloch theorem, because it was previously proven to be valid in some one-dimensional and two-dimensional Hermitian systems, where it is known as the generalization of Bloch's theorem\cite{Viola1,Viola2}. In Section \ref{sec1}, we outline the main steps of our non-Hermitian generalized Bloch theorem for a generic one-dimensional non-Hermitian tight-binding model with two sub-lattice, and provide the extended version to be applicable for arbitrary one-dimensional model. Our main contribution is the derivation of Vieta's formulas for any general polynomials with arbitrary degree, which relate the roots and the coefficients of the characteristic equation. In Section \ref{sec2}, we demonstrate how the non-Hermitian generalized Bloch theorem can help us conveniently analyze the condition of existence or non-existence of non-Hermitian skin effects at a mathematically rigorous level, using two concrete examples including the Su-Schrieffer-Heeger (SSH) chain model with long-range couplings, and a more complicated model called the ladder model\cite{Edvardsson2020}. In Section \ref{sec3}, the non-Hermitian generalized Bloch theorem allows us to explore another extraordinary non-Hermitian phenomena, called the real-space exceptional points\cite{Kunst2019}, whose order scales with the lattice size in the spectrum under the open boundary condition.Additionally, in Section \ref{sec4} we establish the connection between the non-Hermitian generalized Bloch theorem and the generalized Brillouin zone method. Our non-Hermitian generalized Bloch theorem provides an efficient way to analytically study various non-Hermitian phenomena in more general cases. 

\section{Non-Hermitian generalized Bloch theorem}\label{sec1}
To demonstrate the non-Hermitian generalized Bloch theorem for non-Hermitian open systems, we begin with a generic one-dimensional tight-binding lattice model that has two sub-lattice and is protected by sub-lattice symmetry. The second-quantized Hamiltonian $\hat H$ in real space is written as,
\begin{equation}
\eqalign{
\hat{H}= &\sum_{n=1}^N\varepsilon_0(\cd_{n,A}c_{n,A}+\cd_{n,B}c_{n,B})+
t_{0L}\cd_{n,A}c_{n,B}+t_{0R}\cd_{n,B}c_{n,A}\\
&+\sum_{m=1}^{M}\sum_{n=1}^{N-m}\sum_{i=A,B}\sum_{j=A,B}t_{m,R}^{i,j}\cd_{n+m,i}c_{n,j}+t_{m,L}^{i,j}\cd_{n,j}c_{n+m,i},}
\label{H_general}
\end{equation}
where $N$ is the total number of unit cells, and $\cd_{n,A/B}$($c_{n,A/B}$) denotes the creation (annihilation) operator of $A$ or $B$ site on the $n^{\rm th}$ cell, $n=1,..,N$. The first  term in equation~(\ref{H_general}) includes the on-site term with the same energy $\varepsilon_0$ for both sites, and the intra-cell interaction with non-reciprocal hopping parameters $t_{0L},t_{0R}$. The second term in equation~(\ref{H_general}) represents the inter-cell interaction with non-reciprocal hopping parameters $t_{m,R}$ ($t_{m,L}$) along right (left) direction. The parameter $m$ represents the $m^{\rm th}$-nearest cells, and we consider at most $M^{\rm th}$-nearest inter-cell interaction. Translation symmetry is broken in our lattice model due to the open boundary condition, which is defined as the hopping truncation between the $k^{\rm th}$ cell and the $N-M+k,...,N^{\rm th}$ cell, where $k$ ranges from $1$ to $M$.  

The first step in applying the non-Hermitian generalized Bloch theorem is to separate the whole open system into bulk subsystem and boundary subsystem\cite{Viola2}. This allows us to split the eigenvalue equation $\hat{H}|\Psi\rangle = E|\Psi\rangle$ under open boundary condition into the bulk equations and boundary equations. The eigenstates of open system can be written as $|\Psi\rangle = \sum_{n=1}^{N}\phi_{n,A}\cd_{n,A}|0\rangle + \phi_{n,B}\cd_{n,B}|0\rangle$, where $\phi_{n,A/B}$ is the amplitude on the $n^{\rm th}$ cell, and then the bulk equations can be expressed as,
\begin{equation}
\eqalign{
t_{0L}\phi_{n,B}+\sum_{m=1}^{M}(\sum_{j=A,B}t_{m,R}^{A,j}\phi_{n-m,j}+\sum_{i=A,B}t_{m,L}^{i,A}\phi_{n+m,i})=(E-\varepsilon_0)\phi_{n,A},\\
t_{0R}\phi_{n,A}+\sum_{m=1}^{M}(\sum_{j=A,B}t_{m,R}^{B,j}\phi_{n-m,j}+\sum_{i=A,B}t_{m,L}^{i,B}\phi_{n+m,i})=(E-\varepsilon_0)\phi_{n,B},}
\label{bulk_general}
\end{equation}
where $n = M+1,..,N-M$. As a generalization of Bloch states under periodical boundary condition, the eigenstates under open boundary condition can be expressed using the ansatz of generalized Bloch states\cite{Viola2}, written as $\phi_{n,A} = \phi_A z^n$ and $\phi_{n,B} = \phi_B z^n$, where $z\in\mathbb{C}$ and $\phi_{A},\phi_{B}$ are coefficients of $z^n$. Therefore, if the eigenvalue $E\neq \varepsilon_0$, the bulk equations in equation~(\ref{bulk_general}) can be transformed into an equation about $z$, which is known as the characteristic equation\cite{Guo2021},
\begin{equation}
\eqalign{
&(t_{0L}+\sum_{m=1}^{M}t_{m,R}^{A,B}z^{-m}+t_{m,L}^{B,A}z^m)(t_{0R}+\sum_{m=1}^{M}t_{m,R}^{B,A}z^{-m}+t_{m,L}^{A,B}z^m)\\
=&(E-\varepsilon_0-\sum_{m=1}^{M}t_{m,R}^{A,A}z^{-m}+t_{m,L}^{A,A}z^m)(E-\varepsilon_0-\sum_{m=1}^{M}t_{m,R}^{B,B}z^{-m}+t_{m,L}^{B,B}z^m).}
\label{2M_general}
\end{equation}
The characteristic equation can be expanded to a polynomial equation with an even degree $4M$,
\begin{equation}
\eqalign{
\omega_0z^{4M}-\omega_1z^{4M-1}+\omega_2z^{4M-2}-,...,-\omega_{4M-1}z +\omega_{4M}=0,\ (\omega_0\neq0).}
\label{w444}
\end{equation}
The polynomial equation above has $4M$ roots $z_1,z_2,...,z_{4M}$, and the coefficients $\omega_0,\omega_1,...,\omega_{4M}$ are determined by the hopping parameters or eigenvalue $E$. According to the fundamental theorem of algebra, the roots are always real numbers or come in complex conjugate pairs, if all polynomial's coefficients are real numbers. It is worth noting that, compared with the previous works where the characteristic equations are quadratic equations\cite{Guo2021,Edvardsson2022,He2020}, we introduce Vieta's theorem for any general polynomial equation with arbitrary degree\cite{Funkhouser1930}. Vieta's formulas relate the polynomial's coefficients to sums of products of the roots as follows,
\begin{equation}
\eqalign{
&\sum_{i}z_i=z_1+z_2+...+z_{4M} = \frac{\omega_{1}}{\omega_0},\\
&\sum_{i,j\ (i<j)}z_iz_j=z_1z_2+z_1z_3+...+z_{4M-1}z_{4M} = \frac{\omega_{2}}{\omega_0},\\
&\sum_{i,j,k\ (i<j<k)}z_iz_jz_k=z_1z_2z_3+z_1z_2z_4...+z_{4M-2}z_{4M-1}z_{4M} = \frac{\omega_{3}}{\omega_0},\\
&...\\
&z_{1}z_{2}...z_{4M}=\frac{\omega_{4M}}{\omega_0}.
}
\label{WD_general}
\end{equation}
Later, we will demonstrate how the Vieta's formulas described in equation (\ref{WD_general}) can efficiently help us analyze the extraordinary non-Hermitian phenomena, such as the existence or non-existence of non-Hermitian skin effects in Section \ref{sec2}, and the appearance of real-space exceptional points in Section \ref{sec3}.

Here we discuss how the boundary conditions determine the permitted generalized Bloch states. Firstly, the formula for generalized Bloch states should be revised to,
\begin{equation}
\phi_{n,A}=\sum_{i=1}^{4M}\phi_A^{(i)}z_i^n,\ \phi_{n,B}=\sum_{i=1}^{4M}\phi_B^{(i)}z_i^n,
\label{open4M}
\end{equation}
where $\phi_{A}^{(i)},\phi_{B}^{(i)}$ are the coefficients for each component $z_i^n$. The open boundary condition requires that the amplitude $\phi_{n,A/B}$ must vanish outside $1\leq n\leq N$\cite{Lee2019}. Therefore, the boundary equations can be expressed as,
\begin{equation}
\phi_{N+1,A}=0,\ \phi_{0,B}=0.
\label{bound_general}
\end{equation}
By inserting equation (\ref{open4M}), the boundary equations shown in equation~(\ref{bound_general}) become,
\begin{equation}
\sum_{i=1}^{4M}\phi_A^{(i)}z_i^{N+1}=0,\ \sum_{i=1}^{4M}\phi_B^{(i)}=0.
\label{det_general}
\end{equation}
Let's write $\Gamma_i = \phi_B^{(i)}/\phi_A^{(i)}$ as the ratio between the coefficients $\phi_A^{(i)}$ and $\phi_B^{(i)}$. Then, the boundary equations become, 
\begin{equation}
\eqalign{
\pmatrix{
z_1^{N+1} 				&z_2^{N+1} 				&\cdots 				&z_{4M}^{N+1} \\
\Gamma_1 				&\Gamma_2 				&\cdots				&\Gamma_{4M}
}
\pmatrix{
\phi_A^{(1)} 				\\
\phi_A^{(2)}				\\
\vdots				\\
\phi_A^{(4M)}			
}
=\bf{0}.
}
\label{det4M}
\end{equation}
From equation (\ref{det4M}), we deduce that under the thermodynamic limit $N\to\infty$, the series of coefficients $\{\phi_A^{(1)},\phi_A^{(2)},...,\phi_A^{(4M)}\}$ admits a non-trivial solution only if there exists at least one pair of complex conjugate roots $z_1,z_2$, or multiple roots with the same absolute values, and all coefficients $\phi_{A}^{(i)},\phi_{B}^{(i)}$ except for these components must become zero in the formula of the permitted generalized Bloch states. Proof by contradiction is used to support this statement. Consider $z_1,z_2$ are a complex conjugate pair with $z_1 = \ee^{\alpha+\ii\theta}$ and $z_2 = \ee^{\alpha-\ii\theta}$, where $\alpha,\theta\in\mathbb{R}$. Let $z_3=\ee^{\beta}$ with $\beta\in\mathbb{R}$, and assume that $\phi_A^{(1),(2),(3)}\neq 0$, while $\phi_A^{(4),..,(4M)}= 0$. The boundary equations can then be written as,
\begin{equation}
\frac{{\rm e}^{{\rm i}\theta(N+1)}\phi_A^{(1)}+{\rm e}^{-{\rm i}\theta(N+1)}\phi_A^{(2)}}{\Gamma_1\phi_A^{(1)}+\Gamma_2\phi_A^{(2)}}={\rm e}^{(\beta-\alpha)(N+1)}\frac{\phi_A^{(3)}}{\Gamma_3}.
\label{gammaN}
\end{equation}
Under the thermodynamic limit that $N\to\infty$, the left side of equation (\ref{gammaN}) becomes a finite non-zero number. However, as long as $\beta\neq\alpha$, the right side of equation (\ref{gammaN}) must be either zero or infinity, making equation (\ref{gammaN}) impossible for any $\phi_A^{(3)}\neq 0$, which contradicts our assumption. Therefore, for the general case where $\beta\neq\alpha$, the boundary equations can be expressed as,
\begin{equation}
z_1^{N+1}\Gamma_2=z_2^{N+1}\Gamma_1,\ \phi_A^{(1),(2)}\neq 0,\ \phi_A^{(3),..,(4M)}= 0.
\label{nontrivial_general}
\end{equation}
Consequently, the Vieta's formulas of the characteristic equation described in equation (\ref{WD_general}) and the boundary equation shown in equation (\ref{nontrivial_general}) jointly determine the value of $\alpha$ and $\theta$. Benefiting from the introduction of $\alpha$ and $\theta$, it is of great convenience to display the expression of eigenvalues and eigenstates in non-Hermitian open systems.

Our non-Hermitian generalized Bloch theorem can be extended to more complicated one-dimensional models. If each unit cell contains $S$ sites, the characteristic equation will become a polynomial equation with an even degree $2MS$. The Vieta's theorem is still effective, which will allow us to analyze various non-Hermitian phenomena at a mathematically rigorous level. Therefore, our non-Hermitian generalized Bloch theorem is generally applicable for one-dimensional non-Hermitian tight-binding models.

\section{Non-Hermitian skin effects and pseudo-Hermitian symmetry}\label{sec2}
\subsection{SSH chain model with long-range couplings}
To illustrate our main results, we provide two concrete and instructive examples. As shown in figure 1(a), the first example is the SSH chain model with long-range couplings, where the hopping parameters described in equation~(\ref{H_general}) are simplified to $t_{0L}=t_{0R}=t_0$, $t_{1,R}^{A,B}=t_{1R}$, $t_{1,L}^{A,B}=t_{1L}$ and $t_{2,L}^{A,B}=t_{2,R}^{A,B}=t_2$. The real-space Hamiltonian of the SSH chain model under open boundary condition is written as,
\begin{equation}
\eqalign{
\hat{H}= &\sum_{n=1}^N\varepsilon_0(\cd_{n,A}c_{n,A}+\cd_{n,B}c_{n,B})+
t_{0}(\cd_{n,A}c_{n,B}+\cd_{n,B}c_{n,A})\\
&+\sum_{n=1}^{N-1}\tR\cd_{n+1,A}c_{n,B}+\tl\cd_{n,B}c_{n+1,A}
+\sum_{n=1}^{N-2}t_2(\cd_{n+2,A}c_{n,B}+ \cd_{n,B}c_{n+2,A}).
}
\label{H1}
\end{equation}
The bulk equations separated from the eigenvalue equation should be expressed as,
\begin{equation}
\eqalign{
&(t_2z^{-2}+t_{1R}z^{-1}+t_0)\phi_B = (E-\varepsilon_0)\phi_A, \\
&(t_0+t_{1L}z+t_2z^2)\phi_A = (E-\varepsilon_0)\phi_B.
}
\label{sshbulk}
\end{equation}
Therefore, the characteristic equation of the SSH chain model can be expressed as a quartic polynomial equation,
\begin{equation}
\eqalign{
&\omega_0z^{4}-\omega_1z^{3}+\omega_2z^{2}-\omega_3z+\omega_{4}=0,\\
&\omega_0 = t_0t_2,\\
&\omega_1=-(t_0\tl +\tR t_2),\\
&\omega_2=t_0^2+\tl\tR+t_2^2-(E-\varepsilon_0)^2,\\
&\omega_3=-(t_0\tR +\tl t_2),\\
&\omega_4 = t_0t_2.
}
\label{chara1}
\end{equation}
Assuming that $\omega_0\neq 0$, the Vieta's formulas can be used to relate the roots $z_1,z_2,z_3,z_4$ to the polynomial's coefficients as follows,
\begin{equation}
\eqalign{
&z_1+z_2+z_3+z_4 = \frac{\omega_{1}}{\omega_0},\\
&z_1z_2+z_1z_3+z_1z_4+z_2z_3+z_2z_4+z_3z_4 = \frac{\omega_{2}}{\omega_0},\\
&z_1z_2z_3+z_1z_2z_4+z_1z_3z_4+z_2z_3z_4 = \frac{\omega_{3}}{\omega_0},\\
&z_1z_2z_3z_4=\frac{\omega_{4}}{\omega_0}.
}
\label{WD1}
\end{equation}
As proven in Section \ref{sec1}, due to the constraint of open boundary condition, a pair of complex roots $z_1 = \ee^{\alpha+\ii\theta}$ and $z_2 = \ee^{\alpha-\ii\theta}$ must appear in the quartic polynomial equation. While another two roots can be expressed as $z_3 = \ee^{-\alpha+\lambda}$ and $z_4 = \ee^{-\alpha-\lambda}$ with $\lambda\in\mathbb{C}$, because $z_1z_2z_3z_4=1$ in our SSH chain model. We only consider the general case where $|z_3|,|z_4|\neq {\rm e}^{\alpha}$, which implies that $\phi_A^{(1),(2)}\neq 0$ and $\phi_A^{(3),(4)}=0$, as proven in Section \ref{sec1}.

\begin{figure}[t]
\centering\includegraphics[width=0.8\linewidth]{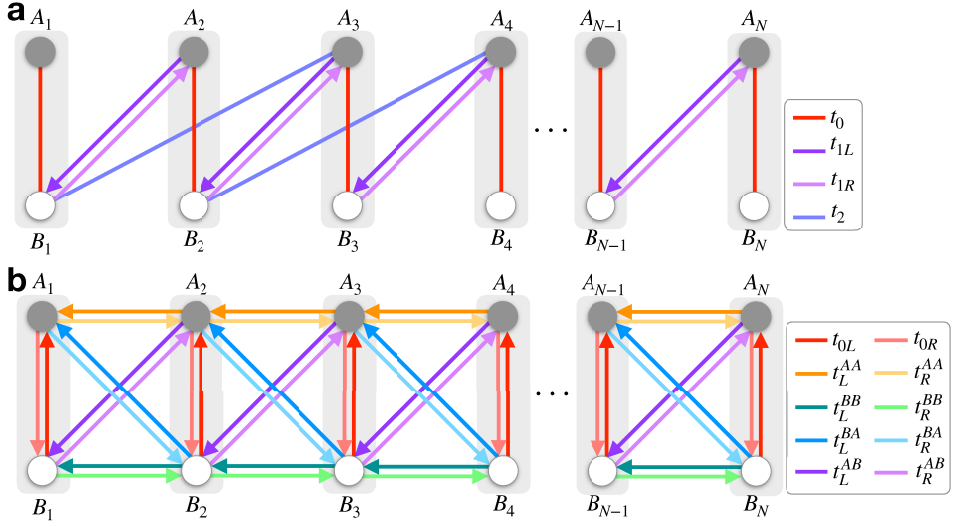}
\caption{
\label{fig1} 
{\bf Schematic views of the one-dimensional non-Hermitian tight-binding models under the open boundary condition.} 
(a) Non-Hermitian SSH chain model with long-range couplings. (b) Ladder model with 5 pairs of non-reciprocal hopping terms.
}
\end{figure}

Next, we discuss how exactly the open boundary condition constrains the permitted generalized Bloch states in our SSH chain model. The ratio between the coefficients $\phi_A^{(i)},\phi_B^{(i)}$ can be solved from the bulk equations in equation (\ref{sshbulk}), 
\begin{equation}
\Gamma_i = \frac{E-\varepsilon_0}{t_0 + t_{1R}z_i^{-1} + t_2 z_j^{-2}},\ i=1,2.
\end{equation}
By substituting the expression of $\Gamma_i$, the boundary equation described in equation (\ref{nontrivial_general}) can be rewritten as,
\begin{equation}
t_0(z_1^{N+1}-z_2^{N+1})+t_{1R}(z_1^N-z_2^N)+t_2(z_1^{N-1}-z_2^{N-1})=0.
\label{tN1}
\end{equation}
Benefiting from the introduction of $\alpha, \theta$, the boundary equation can be further simplified as, 
\begin{equation}
\sin[(N+1)\theta]+\mu_1\sin[N\theta]+\mu_2\sin[(N-1)\theta]=0,
\end{equation}
where $\mu_1={\rm e}^{-\alpha}t_{1R}/t_0$, $\mu_2={\rm e}^{-2\alpha}t_2/t_0$. Although $\theta$ cannot be analytically solved from the boundary equation above, an important fact is discovered that $\theta$ can be regarded as a known variable because its value is continuous within $(0,\pi)$ under the thermodynamic limit $N\to\infty$. 

\begin{figure}[t]
\centering\includegraphics[width=1\linewidth]{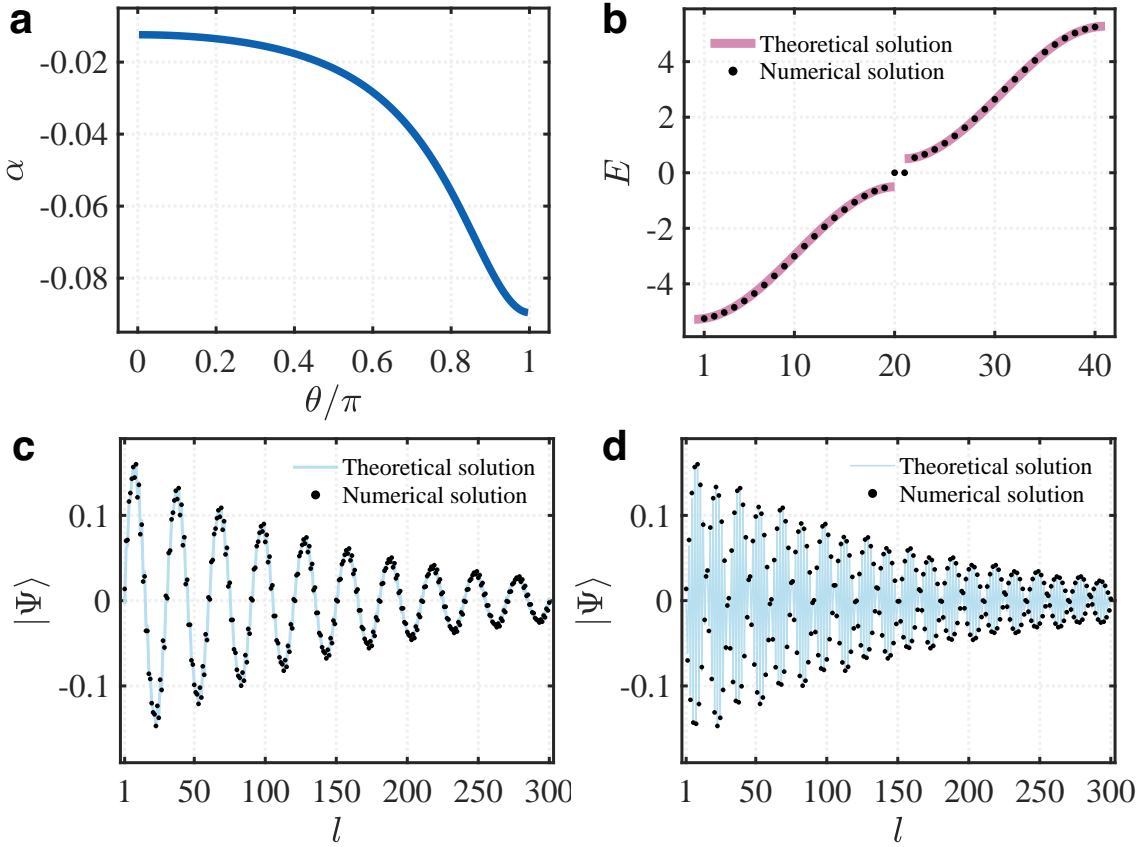}
\caption{\label{fig2} 
{\bf Results of non-Hermitian generalized Bloch theorem for SSH chain model under open boundary condition.} 
(a) Localization parameter $\alpha$ solved from $\cos\theta=(\omega_1\ee^{\alpha}-\omega_3\ee^{-\alpha})/[2\omega_0(\ee^{2\alpha}-\ee^{-2\alpha})],\ \theta\in(0,\pi)$. (b) Eigenvalues $E$ and (c,d) eigenstates $|\Psi\rangle$, where the solid lines represent the theoretical solution derived from our non-Hermitian generalized Bloch theorem, and the dots represent the numerical solution calculated from the diagonalization of Hamiltonian matrix. $l$ represents the $l^{\rm th}$ site in the one-dimensional chain. The hopping parameters are set as $\varepsilon_0 = 0$, $t_0 = 1$, $t_{1L}=2.5$, $t_{1R}=3.5$ and $t_2 = 1.3$.
}
\end{figure}

Through inserting $\alpha,\theta,\lambda$, the Vieta's formulas are derived as,
\begin{equation}
\eqalign{
&2\ee^{\alpha}\cos\theta+2\ee^{-\alpha}\cosh\lambda= \frac{\omega_{1}}{\omega_0},\\
&2\cosh2\alpha+4\cos\theta\cosh\lambda=\frac{\omega_{2}}{\omega_0},\\
&2\ee^{-\alpha}\cos\theta+2\ee^{\alpha}\cosh\lambda = \frac{\omega_{3}}{\omega_0},
}
\label{WDa1}
\end{equation}
where $\lambda$ serves as an intermediate variable during the derivation process. Finally, the parameter $\alpha$ can be expressed analytically in terms of the given variable $\theta$,
\begin{equation}
\cos\theta=\frac{\omega_1\ee^{\alpha}-\omega_3\ee^{-\alpha}}{2\omega_0(\ee^{2\alpha}-\ee^{-2\alpha})},\ \theta\in(0,\pi), \alpha\neq 0,
\label{alph}
\end{equation}
where $\omega_0,\omega_1,\omega_3$ are determined only by the hopping parameters. As shown in figure \ref{fig2}(a), the parameter $\alpha$ is not a constant number, but varies with $\theta\in(0,\pi)$. Furthermore, the solvable eigenvalue $E$ of the SSH chain model can be analytically expressed as a function of $\alpha(\theta)$,
\begin{equation}
E = \varepsilon_0\pm\sqrt{t_0^2+t_{1L}t_{1R}+t_2^2-t_0t_2(2\cosh2\alpha+\frac{2\omega_1\omega_3\cosh2\alpha-\omega_1^2-\omega_3^2}{4\omega_0^2\sinh^22\alpha})}.
\end{equation}
In addition, the generalized Bloch states can be expressed as a function of $\alpha$ and $\theta$. Since $\phi_B^{(1)}=\Gamma_1 \phi_A^{(1)}$, $\phi_B^{(2)}=\Gamma_2 \phi_A^{(2)}$, $\phi_B^{(1)} + \phi_B^{(2)}=0$, and $\phi_A^{(3)},\phi_A^{(4)}=0$, the eigenstates $\phi_{n,A}$ and $\phi_{n,B}$ can be derived as,
\begin{equation}
\eqalign{
\phi_{n,A}& = \phi_A^{(1)}z_1^n+\phi_A^{(2)}z_2^n\\
&= \phi_A^{(1)}(z_1^n-\frac{\Gamma_1}{\Gamma_2}z_2^n)\\
&=\phi_0{\rm e}^{\alpha n}(\sin[n\theta]+
\mu_1\sin[(n-1)\theta]+\mu_2 \sin[(n-2)\theta]),
}
\end{equation}
\begin{equation}
\eqalign{
\phi_{n,B}& = \phi_B^{(1)}z_1^n+\phi_B^{(2)}z_2^n\\
&=\Gamma_1 \phi_A^{(1)}(z_1^n-z_2^n)\\
&=\phi_0(E-\varepsilon_0){\rm e}^{\alpha n}\sin[n\theta],
}
\end{equation}
where $\phi_0=2{\rm i} \phi_A^{(1)}t_0/(t_0 + t_{1R}z_1^{-1} + t_2 z_1^{-2})$, and $\theta\in(0,\pi)$ under the thermodynamic limit. 

The analytical expression of the generalized Bloch states reveals that, the eigenstates with $E\neq \varepsilon_0$ that can be solved under open boundary condition are entirely evolved from the bulk states under periodical boundary condition, but modified by a factor of ${\rm e}^{\alpha n}$ in space, where $\alpha$ is called the localization parameter. If $\alpha\neq 0$, all of the solvable eigenstates are localized at one side of the lattice model, a phenomenon known as non-Hermitian skin effects. When $\alpha < 0$, the skin modes are localized at $n=1$ side with penetration length $-1/\alpha$, while when $\alpha > 0$, the skin modes are localized at $n=N$ side with penetration length $1/\alpha$. To confirm the validity of our non-Hermitian generalized Bloch theorem, we have also calculated the eigenvalues and eigenstates numerically by diagonalizing the Hamiltonian matrix. As shown in figures \ref{fig2}(b,c,d), the numerical solutions and the theoretical solutions agree with each other precisely.

While the non-Hermitian skin effect is a ubiquitous phenomenon in non-Hermitian open systems, we need to focus on a special case where non-Hermitian systems cannot exhibit non-Hermitian skin effects. One such system is the parity-time symmetric system with balanced gain and loss, which can provide relatively stable states in realistic non-Hermitian systems and has the potential to be useful for practical applications\cite{Xu2021}. In this work, our non-Hermitian generalized Bloch theorem is highly effective in analyzing the non-existence of non-Hermitian skin effects. By setting $\alpha=0$, the Vieta's formulas described in equation~(\ref{WDa1}) simplify to,
\begin{equation}
\eqalign{
&2\cos\theta+2\cosh\lambda= \frac{\omega_{1}}{\omega_0},\\
&2+4\cos\theta\cosh\lambda=\frac{\omega_{2}}{\omega_0},\\
&2\cos\theta+2\cosh\lambda = \frac{\omega_{3}}{\omega_0}.
}
\end{equation} 
Thus, the non-Hermitian skin effects vanish only when $\omega_1=\omega_3$, which means either $t_{1L}=t_{1R}$ or $t_0=t_2$. In the former case, the system reduces to a Hermitian one, while in the latter case, it remains non-Hermitian but with $t_0=t_2$. Symmetry plays a crucial role in the absence of non-Hermitian skin effects. When $t_0=t_2$, our SSH chain model is protected by the pseudo-Hermitian symmetry\cite{Mostafazadeh2004,Lieu2018,Yuce2018}, where the total hopping along right direction is equivalent to the total hopping along left direction. It can be described as 
\begin{equation}
H^{\dagger} = \rho H \rho ^{-1},
\end{equation}
where $\rho$ is a Hermitian and invertible matrix. More details on the proof of pseudo-Hermitian symmetry in the SSH chain model are illustrated in \ref{Ap}.   

\subsection{Ladder model with non-reciprocal interaction}
\begin{figure}[t]
\centering\includegraphics[width=1\linewidth]{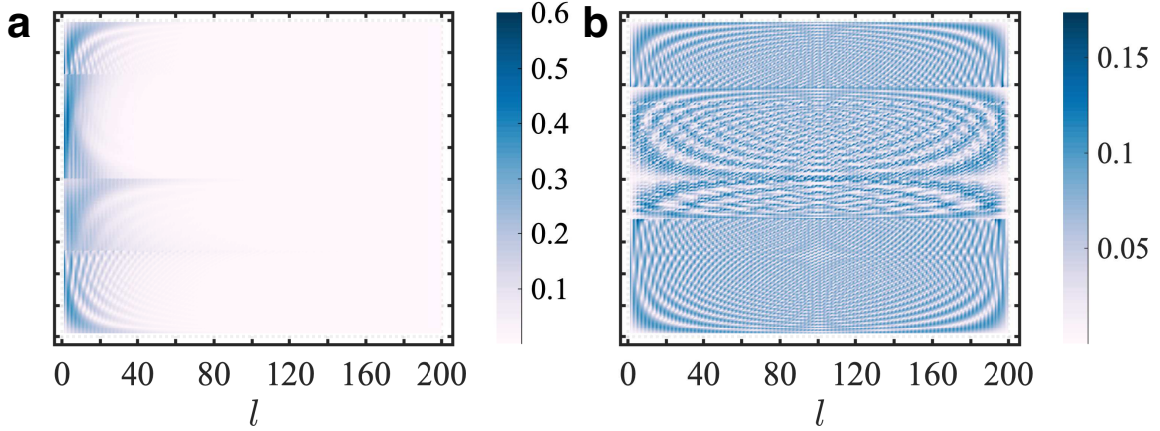}
\caption{
\label{fig3} 
{\bf Absolute values of the eigenstates for the ladder model, by numerically diagonalizing the Hamiltonian matrix under open boundary condition.} (a) Case without pseudo-Hermitian symmetry. (b) Case with pseudo-Hermitian symmetry. $l$ represents the $l$th site of the lattice model. The hopping parameters are set as $\varepsilon_0 = 0$, $t_{0L} = 1$, $t_{0R} = 0.5$, $t_L^{AA}=1.2$, $t_R^{AA}=0.6$, $t_L^{BB}=0.6$, $t_R^{BB}=1.2$, $t_R^{AB}=1$, $t_L^{BA}=3$, $t_R^{BA}=1.5$, $t_L^{AB}=1.1$ in (a) and $t_L^{AB}=0.5$ in (b). 
}
\end{figure}
To demonstrate the general applicability of our non-Hermitian generalized Bloch theorem, a more complicated one-dimensional lattice model is proposed. As shown in figure 1(b), the second example is a modification of Lee's model\cite{Lee2016}, known as the ladder model\cite{Edvardsson2020}. In this model, each site is coupled with five neighboring sites, and the hopping terms between sites are non-reciprocal, resulting in ten degrees of freedom. The condition of the non-existence of non-Hermitian skin effects in the ladder model, is far more complex than that of SSH chain model. Fortunately, our non-Hermitian generalized Bloch theorem provides a convenient solution to this problem.

The characteristic equation of the ladder model can be expressed as a quartic polynomial equation,
\begin{equation}
\eqalign{
&\omega_0z^{4}-\omega_1z^{3}+\omega_2z^{2}-\omega_3z+\omega_{4}=0, \ (\omega_0\neq 0)
}
\label{chara2}
\end{equation}
where
\begin{equation}
\eqalign{
&\omega_0 = t_L^{AA}t_L^{BB}-t_L^{AB}t_L^{BA},\\
&\omega_1=(E-\varepsilon_0)(t_L^{AA}+t_L^{BB})+t_{0L}t_L^{AB}+t_{0R}t_L^{BA},\\
&\omega_2=(E-\varepsilon_0)^2 + t_L^{AA}t_R^{BB}+t_L^{BB}t_R^{AA}-t_{0L}t_{0R}-t_L^{AB}t_R^{AB}-t_L^{BA}t_R^{BA},\\
&\omega_3=(E-\varepsilon_0)(t_R^{AA}+t_R^{BB})+t_{0L}t_R^{BA}+t_{0R}t_R^{AB},\\
&\omega_4 = t_R^{AA}t_R^{BB}-t_R^{AB}t_R^{BA}.
}
\label{chara2}
\end{equation}
By substituting $z_1,z_2,z_3,z_4$ with $z_1 = \ee^{\alpha+\ii\theta}$, $z_2 = \ee^{\alpha-\ii\theta}$, $z_3 = \ee^{\beta+\lambda}$ and $z_4 = \ee^{\beta-\lambda}$, where $\alpha,\beta,\theta\in\mathbb{R}$ and $\lambda\in\mathbb{C}$, the Vieta's formulas for the ladder model can be written as,
\begin{equation}
\eqalign{
&2\ee^{\alpha}\cos\theta + 2\ee^{\beta}\cosh\lambda  = \frac{\omega_{1}}{\omega_0},\\
&\ee^{2\alpha}+\ee^{2\beta}+4\ee^{\alpha+\beta}\cos\theta\cosh\lambda = \frac{\omega_{2}}{\omega_0},\\
&2\ee^{\alpha+2\beta}\cos\theta + 2\ee^{2\alpha+\beta}\cosh\lambda = \frac{\omega_{3}}{\omega_0},\\
&\ee^{2\alpha+2\beta} = \frac{\omega_{4}}{\omega_0}.
}
\end{equation}
By setting $\alpha=0$, the Vieta's formulas are further derived as
\begin{equation}
\eqalign{
&2\cos\theta + 2\ee^{\beta}\cosh\lambda  = \frac{\omega_{1}}{\omega_0},\\
&1+\ee^{2\beta}+4\ee^{\beta}\cos\theta\cosh\lambda = \frac{\omega_{2}}{\omega_0},\\
&2\ee^{2\beta}\cos\theta + 2\ee^{\beta}\cosh\lambda = \frac{\omega_{3}}{\omega_0},\\
&\ee^{2\beta} = \frac{\omega_{4}}{\omega_0},
}
\end{equation}
where $\beta,\lambda$ serve as the intermediate variables. Therefore, the sufficient and necessary condition of the non-existence of non-Hermitian skin effects is expressed as,
\begin{equation}
\forall \theta, 2(\omega_0-\omega_4)\cos\theta=\omega_1-\omega_3.
\end{equation}
Consequently, the non-Hermitian skin effects disappear only if $\omega_0=\omega_4$ and $\omega_1=\omega_3$, which means,
\begin{equation}
\label{cases}
\cases{t_L^{AA}t_L^{BB}-t_L^{AB}t_L^{BA}=t_R^{AA}t_R^{BB}-t_R^{AB}t_R^{BA}&\\
		 t_L^{AA}+t_L^{BB}=t_R^{AA}+t_R^{BB}&\\
		 t_{0L}t_L^{AB}+t_{0R}t_L^{BA}=t_{0L}t_R^{BA}+t_{0R}t_R^{AB}&\\}.
\end{equation}

As shown in figure \ref{fig3}(a), when $t_L^{AB}=1.1$, all the eigenstates are localized at the left side of lattice because the condition stated in equation (\ref{cases}) is not satisfied. Conversely, in figure \ref{fig3}(b) where $t_L^{AB}=0.5$, the eigenstates behaves like the bulk states without localization, which satisfies the condition in equation (31) and ensures the disappearance of non-Hermitian skin effects. Our non-Hermitian generalized Bloch theorem enables efficient analysis of non-Hermitian skin effects in a broader range of cases.
 

\section{Real-space exceptional points}\label{sec3}
In this section, we employ our non-Hermitian generalized Bloch theorem to investigate the real-space exceptional points (EPs), a phenomenon characterized by the coalescence of both eigenvalues and eigenstates under open boundary conditions. Unlike Bloch EPs, whose order are restricted by the dimensionality of the Bloch Hamiltonian, the order of real-space EPs increases with the size of the lattice\cite{Kunst2019}. Consequently, higher-order EPs (greater than second order) can be achieved under open boundary conditions, leading to a significant enhancement of sensitivity in non-Hermitian systems\cite{Budich2020,Hodaei2017}. To harness these advantages for further practical applications, the full understanding and control of real-space EPs is of great significance. 

\begin{figure}[t]
\centering\includegraphics[width=1\linewidth]{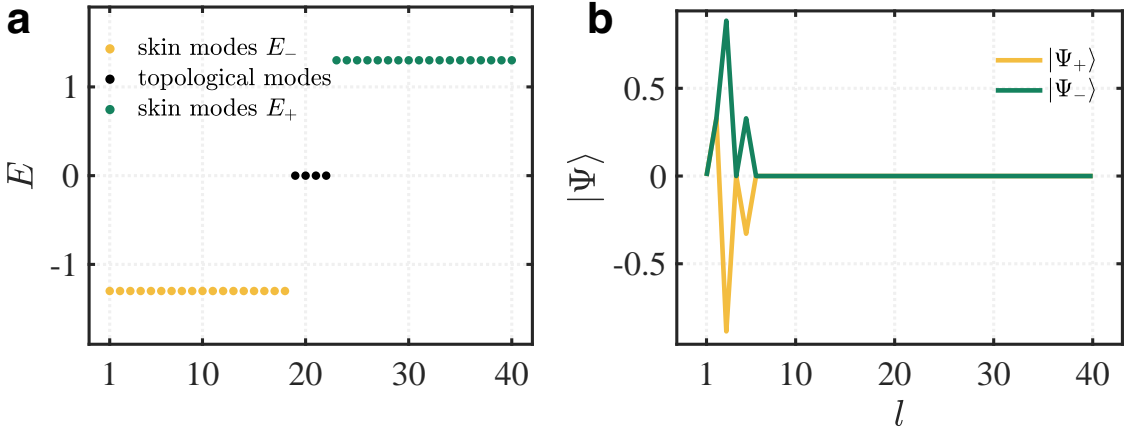}
\caption{
\label{fig4} 
{\bf Real-space exceptional points for SSH chain model under open boundary condition. } Analytical solution of eigenvalues in (a), and eigenstates in (b). $l$ represents the $l$th site of the lattice model. The hopping parameters are set as $\varepsilon_0 =0$, $t_0 = 0$, $t_{1L} = 0$, $t_{1R}=3.5$, and $t_2 = 1.3$.
}
\end{figure}

Previous works have revealed that, real-space EPs occur when the system has a completely unidirectional hopping, so that all states are piled up at one end of the system\cite{Kunst2019}. Based on this insight, we can conclude that the real-space EPs appear when the localization parameter satisfies $\alpha=\mp\infty$, which ensures that the eigenstates are strongly accumulated at either the $n=1$ end or $n=N$ end. For the SSH chain model, if we set $\alpha=-\infty$ and assume that $\theta$ is a finite number, the first equation of the Vieta's formulas described in equation~(\ref{WDa1}) is transformed as,
\begin{equation}
\eqalign{
-\frac{t_0\tl +\tR t_2}{t_0t_2}=\infty.}
\label{Oz0}
\end{equation}
To make equation (\ref{Oz0}) possible, we can either set $t_{1L}=0,t_0=0$ or set $t_{1R}=0,t_2 =0$. Fortunately, we can confirm these results by analytically diagonalizing the Hamiltonian matrix. Figure \ref{fig4} shows the case where $t_{1L}=0$ and $t_0=0$, demonstrating that there exists a pair of real-space EPs with order $N-2$. Their eigenvalues and eigenstates are analytically expressed as,
\begin{equation}
E_{\pm}=\varepsilon_0\pm t_2,\ |\Psi_{\pm}\rangle=[0,\pm 1,t_{1R}/t_2,0,1,0,0,...,0]^{\rm T}. 
\end{equation} 
The eigenstates are strongly accumulated at the $n=1$ end,  because the hopping terms along right direction are completely truncated. Similarly, if we set $\alpha=\infty$ and assume $\theta$ is a finite number, the third equation of the Vieta's formulas is transformed as, 
\begin{equation}
\eqalign{
-\frac{t_0\tR +\tl t_2}{t_0t_2}=\infty.
}
\label{Oz1}
\end{equation}
There are two possible solutions that $t_{1L}=0,t_2=0$ or $t_{1R}=0,t_0=0$. For example, when we set $t_{1R}=0$ and $t_0=0$, a pair of real space EPs with order $N$ exists, and their eigenstates are strongly accumulated at the $n=N$ end because the hopping terms along left direction are truncated. Therefore, a total of four cases of real-space EPs have been discovered in the SSH chain model through the non-Hermitian generalized Bloch theorem, as shown in Table 1. 

\begin{table}[t]
\caption{\label{tabone}Real-space exceptional points for the SSH chain model and the ladder model.} 
\begin{indented}
\lineup
\item[]\begin{tabular}{@{}*{5}{l}}
\br                              
Model&Parameters &Order & Eigenvalues &Eigenstates \cr 
\mr
      & $t_0=0,t_{1L}=0$ & $N-2$ & $\varepsilon_0\pm t_2$ & $[0,\pm 1,t_{1R}/t_2,0,1,0,0,...,0]^{\rm T}$ \cr
SSH   & $t_0=0,t_{1R}=0$ & $N-2$ & $\varepsilon_0\pm t_2$ & $[0,...,0,0,1,0,t_{1L}/t_2,\pm 1,0]^{\rm T}$ \cr 
chain & $t_2=0,t_{1L}=0$ & $N$   & $\varepsilon_0\pm t_0$ & $[0,...,0,0,\pm1,1]^{\rm T}$ \cr 
      & $t_2=0,t_{1R}=0$ & $N$   & $\varepsilon_0\pm t_0$ & $[1,\pm 1,0,0,...,0]^{\rm T}$ \cr 
      \mr
      & $t_L^{AA},t_L^{BB},t_L^{AB},t_L^{BA}=0$ & $N$ & $\varepsilon_0\pm \sqrt{t_{0L}t_{0R}}$ & $[0,...,0,0,\pm \frac{\sqrt{t_{0L}t_{0R}}}{t_{0R}},1]^{\rm T}$ \cr
      
& $t_R^{AA},t_R^{BB},t_R^{AB},t_R^{BA}=0$ & $N$ & $\varepsilon_0\pm \sqrt{t_{0L}t_{0R}}$ & $[\frac{\sqrt{t_{0L}t_{0R}}}{t_{0R}},\pm 1,0,0,...,0]^{\rm T}$ \cr 
ladder & $t_L^{AA}t_L^{BB}-t_L^{AB}t_L^{BA}=0,$ && \cr
 model & $t_L^{AA}+t_L^{BB}=0, t_{0L},t_{0R}=0$ & $N-1$   & $\varepsilon_0\pm \sqrt{E_L}$ & $[0,...,0,0,\pm P_L,\pm Q_L,S_L,1]^{\rm T}$ \cr 
      & $t_R^{AA}t_R^{BB}-t_R^{AB}t_R^{BA}=0,$ && \cr
 & $t_R^{AA}+t_R^{BB}=0, t_{0L},t_{0R}=0$ & $N-1$   & $\varepsilon_0\pm \sqrt{E_R}$ & $[\pm P_R,\pm Q_R,S_R,1,0,0,...,0]^{\rm T}$ \cr   
\br
\end{tabular}
\end{indented}
\end{table}

The non-Hermitian generalized Bloch theorem is also applicable for the ladder model. As shown in Table 1, real-space EPs can occur at the point where $t_L^{AA}=t_L^{BB}=t_L^{AB}=t_L^{BA}=0$ or $t_R^{AA}=t_R^{BB}=t_R^{AB}=t_R^{BA}=0$, since every inter-cell hopping term along left/right direction vanishes. It is worth noting that, the condition for real-space EPs in the ladder model is not as strict as that in SSH chain model. For example, assuming that $t_L^{AA},t_L^{BB},t_L^{AB},t_L^{BA}\neq 0$, if we choose $t_L^{AA}t_L^{BB}-t_L^{AB}t_L^{BA}=0,t_L^{AA}+t_L^{BB}=0, t_{0L}=t_{0R}=0$ to ensure the coefficients $\omega_0 = 0$ and $\omega_1=0$, then a pair of real-space EPs appear with order $N-1$, and their eigenvalues and eigenstates are analytically expressed as,
\begin{equation}
\eqalign{
& E_{\pm}=\varepsilon_0\pm \sqrt{E_L},\\
& |\Psi_{\pm}\rangle=[0,...,0,0,\pm P_L,\pm Q_L,S_L,1]^{\rm T},}
\end{equation}
where 
\begin{equation}
\eqalign{
& E_L=[t_L^{AA}t_L^{AB}(t_R^{AA}-t_R^{BB})+{t_L^{AB}}^2t_R^{AB}-{t_L^{AA}}^2t_R^{BA}]/t_L^{AB},\\
& P_L=\sqrt{E_L}t_L^{AA}/(t_L^{AA}t_R^{BA}+t_L^{AB}t_R^{BB}),\\
& Q_L=\sqrt{E_L}t_L^{AB}/(t_L^{AA}t_R^{BA}+t_L^{AB}t_R^{BB}),\\
& S_L=(t_L^{AA}t_R^{AA}+t_L^{AB}t_R^{ABB})/(t_L^{AA}t_R^{BA}+t_L^{AB}t_R^{BB}).
}
\end{equation}
 Although the inter-cell hopping terms $t_L^{AA},t_L^{BB},t_L^{AB},t_L^{BA}$ along left direction still survive, their compensation of each other prevents the  hopping between the nearest cells, leading to the strong accumulation of eigenstates. Another case of real-space EP occurs when $t_R^{AA}t_R^{BB}-t_R^{AB}t_R^{BA}=0,t_R^{AA}+t_R^{BB}=0, t_{0L}=t_{0R}=0$. In summary, the non-Hermitian Bloch theorem allows us to discover more cases of real-space EPs in our non-Hermitian models, providing an effective way to fully understand and control non-Hermitian open systems.

\section{Connection with the generalized Brillouin zone method}\label{sec4}
In this section, we establish the connection between our non-Hermitian generalized Bloch theorem and the well-known generalized Brillouin zone(GBZ) method\cite{Yao2018,Yokomizo2019}, which is used to construct the non-Hermitian bulk-boundary correspondence. The GBZ method leads to the same conclusions regarding the characteristic equation and the open boundary condition as our non-Hermitian generalized Bloch theorem. However, in the GBZ method, the eigenvalue $E$ in the characteristic equation is considered as a given variable, thus the roots of the characteristic equation in equation (\ref{w444}) can be numerically solved, which satisfy $|z_1|\leq|z_2|\leq...|z_{2M}|\leq|z_{2M+1}|\leq...\leq|z_{4M}|$. To get the continuum bands that reproduce the band structure for a large crystal under the open boundary condition, there must be,
\begin{equation}
\eqalign{
|z_{2M}|=|z_{2M+1}|.
}
\end{equation}
The trajectory of $z_{2M}$ and $z_{2M+1}$ gives the generalized Brillouin zone $\mathcal{C}_z$, which determines the continuum bands\cite{Yokomizo2019}.

\begin{figure}[t]
\centering\includegraphics[width=0.55\linewidth]{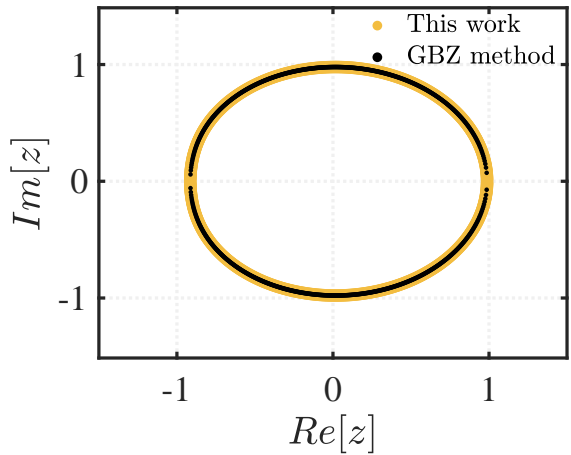}
\caption{
\label{fig5}
{\bf The loop of generalized Brillouin zone in the complex plane for SSH chain model.} The yellow dots denote the results solved using our non-Hermitian generalized Bloch theorem, where $z = {\rm e}^{\alpha\pm{\rm i}\theta},\theta\in (0,\pi)$, while the black dots denote the results calculated from the generalized Brillouin zone (GBZ) method. The hopping parameters are set as $\varepsilon_0 = 0$, $t_0 = 1$, $t_{1L}=2.5$, $t_{1R}=3.5$ and $t_2 = 1.3$.}
\end{figure}

Therefore, we can establish a connection between our non-Hermitian generalized Bloch theorem and the GBZ method by relating the localization parameter $\alpha(\theta)$ to the generalized Brillouin zone $\mathcal{C}_z$. Their relationship can be written as,
\begin{equation}
\alpha = \log |\mathcal{C}_z|.
\end{equation} 
As shown in figure \ref{fig5}, for the SSH chain model with long-range couplings, the trajectory of the generalized Brillouin zone $\mathcal{C}_z$ calculated from the GBZ method and solved using our non-Hermitian generalized Bloch theorem are in precise agreement with each other. The loop of $\mathcal{C}_z$ is not a unit circle in the complex plane, indicating that $\alpha\neq 0$ and the non-Hermitian skin effects can occur. An advantage of our method is that we do not require any prior information about the spectrum under the open boundary condition, and the localization parameter $\alpha$ can be expressed analytically in terms of the given $\theta$, as described by equation (\ref{alph}). 


\section{Conclusions}
In summary, we extend the existing exact solution approach to make it applicable for general one-dimensional non-Hermitian models by introducing the Vieta's formulas for any general polynomial equations. Our non-Hermitian generalized Bloch theorem provides a convenient way to study extraordinary non-Hermitian phenomena. By setting $\alpha=0$ in the Vieta's formulas, our method enables us to explore the disappearance of non-Hermitian skin effects at a mathematically rigorous level, where pseudo-Hermitian symmetry plays a significant role. By setting $\alpha=\mp\infty$ in the Vieta's formulas, our method allows us to analyze the real-space exceptional point, and we discover four cases of exceptional points in the SSH chain model and ladder model, respectively. Furthermore, we establish the connection between our non-Hermitian generalized Bloch theorem and the generalized Brillouin zone method through the relationship $\alpha=\log |\mathcal{C}_z|$, and provide the analytical expression of the generalized Brillouin zone in more general cases. However, there are still some open questions to be addressed in future. For example, we need to further study a special case where more than two roots have the same absolute value in a one-dimensional chain model. Moreover, our method has the potential to be extended to higher-dimensional non-Hermitian systems.

\ack{The authors thank for the support from the National Natural Science Foundation of China (Grant Nos. 60578047, 61427815), and the Natural Science Foundation of Shanghai (Grant Nos. 17ZR1402200, 13ZR1402600).}

\section*{Data availability statement}
The data that support the plots within this paper are available from the corresponding author on reasonable request.

\appendix
\section{Proof of pseudo-Hermitian symmetry}\label{Ap}
Here, we give a proof that when $t_0$ = $t_2$, our SSH chain model is protected by the pseudo-Hermitian symmetry. Assuming that $\varepsilon_0 = 0$ and the hopping parameters are all real numbers, when $t_0=t_2$, the Bloch Hamiltonian becomes
\begin{equation}
\eqalign{
H(k) & =\pmatrix{
0 &h_a(k)\\
h_b(k) & 0
},\\
h_a(k) &= {\rm e}^{-{\rm i} k}(t_{1R} + 2t_0\cos k),\\ 
h_b(k) &= {\rm e}^{{\rm i} k}(t_{1L} + 2t_0\cos k).
}
\end{equation}
A canonical form of the operator $\rho$ and its inverse should be constructed from the right eigenvector and left eigenvector\cite{Mostafazadeh2004}, derived as
\begin{equation}
\eqalign{
\rho &= |\psi_1^{L}\rangle\langle\psi_1^{ L}| +|\psi_2^{ L}\rangle\langle\psi_2^{L}| = \pmatrix{
|\frac{\sqrt{h_a(k)h_b(k)}}{h_a(k)}|^2 &0\\
0&1
},\\
\rho^{-1} &= |\psi_1^{ R}\rangle\langle\psi_1^{R}| +|\psi_2^{ R}\rangle\langle\psi_2^{ R}| = \pmatrix{
|\frac{\sqrt{h_a(k)h_b(k)}}{h_b(k)}|^2 &0\\
0&1
}.
}
\end{equation} 
Therefore, $H^{\dagger}(k)=\rho H(k)\rho^{-1}$ requires that $h_a^*(k)h_b^*(k)=|h_a(k)h_b(k)|$. Obviously, when $h_a(k)h_b(k)\geq 0$, the Bloch Hamiltonian has purely real spectrum, which is the so-called pseudo-Hermitian unbroken phase.  

\section*{References}

\begin{thebibliography}{10}
\expandafter\ifx\csname url\endcsname\relax
  \def\url#1{{\tt #1}}\fi
\expandafter\ifx\csname urlprefix\endcsname\relax\def\urlprefix{URL }\fi
\providecommand{\eprint}[2][]{\url{#2}}

\bibitem{Bender1998}
Bender C~M and Boettcher S 1998 {\em Phys. Rev. Lett.\/} {\bf 80}(24)
  5243--5246

\bibitem{El-Ganainy2018}
El-Ganainy R, Makris K~G, Khajavikhan M, Musslimani Z~H, Rotter S and
  Christodoulides D~N 2018 {\em Nat. Phys.\/} {\bf 14} 11--19

\bibitem{Mostafazadeh2002}
Mostafazadeh A 2002 {\em J. Math. Phys.\/} {\bf 43} 205--214

\bibitem{Mostafazadeh2002b}
Mostafazadeh A 2002 {\em J. Math. Phys.\/} {\bf 43} 2814--2816

\bibitem{Feshbach1958}
Feshbach H 1958 {\em Ann. Phys.\/} {\bf 5} 357--390 ISSN 0003-4916

\bibitem{Lindblad1976}
Lindblad G 1976 {\em Commun. Math. Phys.\/} {\bf 48} 119 -- 130

\bibitem{Colbert1992}
Colbert D~T and Miller W~H 1992 {\em J. Chem. Phys.\/} {\bf 96} 1982--1991

\bibitem{Zhu2020}
Zhu X, Wang H, Gupta S~K, Zhang H, Xie B, Lu M and Chen Y 2020 {\em Phys. Rev.
  Research\/} {\bf 2}(1) 013280

\bibitem{Zhong2021}
Zhong J, Wang K, Park Y, Asadchy V, Wojcik C~C, Dutt A and Fan S 2021 {\em
  Phys. Rev. B\/} {\bf 104}(12) 125416

\bibitem{Chen2021}
Zhang X, Tian Y, Jiang J~H, Lu M~H and Chen Y~F 2021 {\em Nat. Commun.\/} {\bf
  12} 5377

\bibitem{Wang2021}
Wang K, Dutt A, Yang K~Y, Wojcik C~C, Vučković J and Fan S 2021 {\em
  Science\/} {\bf 371} 1240--1245

\bibitem{Parto2021}
Parto M, Liu Y~G~N, Bahari B, Khajavikhan M and Christodoulides D~N 2021 {\em
  Nanophotonics\/} {\bf 10} 403--423

\bibitem{Yoshida2020}
Yoshida T, Mizoguchi T and Hatsugai Y 2020 {\em Phys. Rev. Research\/} {\bf
  2}(2) 022062

\bibitem{Hofmann2020}
Hofmann T, Helbig T, Schindler F, Salgo N, Brzezi\ifmmode~\acute{n}\else
  \'{n}\fi{}ska M, Greiter M, Kiessling T, Wolf D, Vollhardt A,
  Kaba\ifmmode~\check{s}\else \v{s}\fi{}i A, Lee C~H, Bilu\ifmmode
  \check{s}\else \v{s}\fi{}i\ifmmode~\acute{c}\else \'{c}\fi{} A, Thomale R and
  Neupert T 2020 {\em Phys. Rev. Research\/} {\bf 2}(2) 023265

\bibitem{Xu2021}
Xu K, Zhang X, Luo K, Yu R, Li D and Zhang H 2021 {\em Phys. Rev. B\/} {\bf
  103}(12) 125411

\bibitem{Zhang2021}
Zhang X, Xu K, Liu C, Song X, Hou B, Yu R, Zhang H, Li D and Li J 2021 {\em
  Commun. Phys.\/} {\bf 4} 166

\bibitem{Zou2021}
Zou D, Chen T, He W, Bao J, Lee C~H, Sun H and Zhang X 2021 {\em Nat.
  Commun.\/} {\bf 12} 7201

\bibitem{Kononchuk2022}
Kononchuk R, Cai J, Ellis F, Thevamaran R and Kottos T 2022 {\em Nature\/} {\bf
  607} 697--702 ISSN 1476-4687

\bibitem{Jing2018}
Zhu W, Fang X, Li D, Sun Y, Li Y, Jing Y and Chen H 2018 {\em Phys. Rev.
  Lett.\/} {\bf 121}(12) 124501

\bibitem{Acoustics2021}
Zhang L, Yang Y, Ge Y, Guan Y~J, Chen Q, Yan Q, Chen F, Xi R, Li Y, Jia D, Yuan
  S~Q, Sun H~X, Chen H and Zhang B 2021 {\em Nat. Commun.\/} {\bf 12} 6297

\bibitem{Flebus2020}
Flebus B, Duine R~A and Hurst H~M 2020 {\em Phys. Rev. B\/} {\bf 102}(18)
  180408

\bibitem{Deng2022}
Deng K and Flebus B 2022 {\em Phys. Rev. B\/} {\bf 105}(18) L180406

\bibitem{Heiss2004}
Heiss W~D 2004 {\em J. Phys. A: Math. Gen.\/} {\bf 37} 2455--2464

\bibitem{Bender2007}
Bender C~M 2007 {\em Rep. Prog. Phys.\/} {\bf 70} 947--1018

\bibitem{Chen2017}
Chen W, Kaya~Özdemir S, Zhao G, Wiersig J and Yang L 2017 {\em Nature\/} {\bf
  548} 192--196 ISSN 1476-4687

\bibitem{Gong2018}
Gong Z, Ashida Y, Kawabata K, Takasan K, Higashikawa S and Ueda M 2018 {\em
  Phys. Rev. X\/} {\bf 8}(3) 031079

\bibitem{Ghatak2019}
Ghatak A and Das T 2019 {\em J. Phys.: Condens. Matter\/} {\bf 31} 263001

\bibitem{Ashida2020}
Ashida Y, Gong Z and Ueda M 2020 {\em Adv. Phys.\/} {\bf 69} 249--435

\bibitem{Bergholtz2021}
Bergholtz E~J, Budich J~C and Kunst F~K 2021 {\em Rev. Mod. Phys.\/} {\bf
  93}(1) 015005

\bibitem{Klitzing1980}
Klitzing K~v, Dorda G and Pepper M 1980 {\em Phys. Rev. Lett.\/} {\bf 45}(6)
  494--497

\bibitem{Thouless1982}
Thouless D~J, Kohmoto M, Nightingale M~P and den Nijs M 1982 {\em Phys. Rev.
  Lett.\/} {\bf 49}(6) 405--408

\bibitem{Hatsugai1993}
Hatsugai Y 1993 {\em Phys. Rev. Lett.\/} {\bf 71}(22) 3697--3700

\bibitem{Bernevig2006}
Bernevig B~A, Hughes T~L and Zhang S~C 2006 {\em Science\/} {\bf 314}
  1757--1761

\bibitem{Schnyder2008}
Schnyder A~P, Ryu S, Furusaki A and Ludwig A~W~W 2008 {\em Phys. Rev. B\/} {\bf
  78}(19) 195125

\bibitem{Lee2016}
Lee T~E 2016 {\em Phys. Rev. Lett.\/} {\bf 116}(13) 133903

\bibitem{Yao2018}
Yao S and Wang Z 2018 {\em Phys. Rev. Lett.\/} {\bf 121}(8) 086803

\bibitem{Alvarez2018}
Martinez~Alvarez V~M, Barrios~Vargas J~E and Foa~Torres L~E~F 2018 {\em Phys.
  Rev. B\/} {\bf 97}(12) 121401

\bibitem{Okuma2020}
Okuma N, Kawabata K, Shiozaki K and Sato M 2020 {\em Phys. Rev. Lett.\/} {\bf
  124}(8) 086801

\bibitem{Yokomizo2021}
Yokomizo K and Murakami S 2021 {\em Phys. Rev. B\/} {\bf 104}(16) 165117

\bibitem{Kim2021}
Kim K~M and Park M~J 2021 {\em Phys. Rev. B\/} {\bf 104}(12) L121101

\bibitem{Brody2013}
Brody D~C 2013 {\em J. Phys. A: Math. Theor.\/} {\bf 47} 035305

\bibitem{Yokomizo2019}
Yokomizo K and Murakami S 2019 {\em Phys. Rev. Lett.\/} {\bf 123}(6) 066404

\bibitem{Kawabata2020}
Kawabata K, Okuma N and Sato M 2020 {\em Phys. Rev. B\/} {\bf 101}(19) 195147

\bibitem{Yang2020}
Yang Z, Zhang K, Fang C and Hu J 2020 {\em Phys. Rev. Lett.\/} {\bf 125}(22)
  226402

\bibitem{Kunst2019}
Kunst F~K and Dwivedi V 2019 {\em Phys. Rev. B\/} {\bf 99}(24) 245116

\bibitem{Kunst2018}
Kunst F~K, Edvardsson E, Budich J~C and Bergholtz E~J 2018 {\em Phys. Rev.
  Lett.\/} {\bf 121}(2) 026808

\bibitem{Edvardsson2020}
Edvardsson E, Kunst F~K, Yoshida T and Bergholtz E~J 2020 {\em Phys. Rev.
  Research\/} {\bf 2}(4) 043046

\bibitem{Guo2021}
Guo C~X, Liu C~H, Zhao X~M, Liu Y and Chen S 2021 {\em Phys. Rev. Lett.\/} {\bf
  127}(11) 116801

\bibitem{Guo2021b}
Guo G~F, Bao X~X and Tan L 2021 {\em New J. Phys.\/} {\bf 23} 123007

\bibitem{Liu2021}
Liu Y, Zeng Y, Li L and Chen S 2021 {\em Phys. Rev. B\/} {\bf 104}(8) 085401

\bibitem{Edvardsson2022}
Edvardsson E and Ardonne E 2022 {\em Phys. Rev. B\/} {\bf 106}(11) 115107

\bibitem{Viola1}
Alase A, Cobanera E, Ortiz G and Viola L 2016 {\em Phys. Rev. Lett.\/} {\bf
  117}(7) 076804

\bibitem{Viola2}
Alase A, Cobanera E, Ortiz G and Viola L 2017 {\em Phys. Rev. B\/} {\bf 96}(19)
  195133

\bibitem{He2020}
He Y and Chien C~C 2020 {\em J. Phys.: Condens. Matter\/} {\bf 33} 085501

\bibitem{Funkhouser1930}
Funkhouser H~G 1930 {\em The American Mathematical Monthly\/} {\bf 37} 357--365

\bibitem{Lee2019}
Lee C~H and Thomale R 2019 {\em Phys. Rev. B\/} {\bf 99}(20) 201103

\bibitem{Mostafazadeh2004}
Mostafazadeh A and Batal A 2004 {\em J. Phys. A: Math. Gen.\/} {\bf 37}
  11645--11679

\bibitem{Lieu2018}
Lieu S 2018 {\em Phys. Rev. B\/} {\bf 97}(4) 045106

\bibitem{Yuce2018}
Yuce C and Oztas Z 2018 {\em Sci. Rep.\/} {\bf 8} 17416 ISSN 2045-2322

\bibitem{Budich2020}
Budich J~C and Bergholtz E~J 2020 {\em Phys. Rev. Lett.\/} {\bf 125}(18) 180403

\bibitem{Hodaei2017}
Hodaei H, Hassan A~U, Wittek S, Garcia-Gracia H, El-Ganainy R, Christodoulides
  D~N and Khajavikhan M 2017 {\em Nature\/} {\bf 548} 187--191 ISSN 1476-4687

\end{thebibliography}
\providecommand{\noopsort}[1]{}\providecommand{\singleletter}[1]{#1}%
\providecommand{\newblock}{}

\end{document}